\documentclass[pra,superscriptaddress,twocolumn]{revtex4}

\usepackage{enumerate}
\usepackage{amsfonts,amssymb,amsmath}
\usepackage[]{graphics,graphicx,epsfig}
\usepackage{amsthm}

\usepackage{graphicx}
\usepackage{dcolumn}
\usepackage{natbib}
\usepackage{color}
\usepackage{multirow}

\def\identity{\leavevmode\hbox{\small1\kern-3.8pt\normalsize1}}

\newtheorem{propo}{Proposition}
\newcommand{\be}{\begin{eqnarray}}
\newcommand{\ee}{\end{eqnarray}}
\newcommand{\bpr}{\begin{propo}}
\newcommand{\epr}{\end{propo}}
\newcommand{\bpf}{\begin{proof}}
\newcommand{\epf}{\end{proof}}

\renewcommand{\epsilon}{\varepsilon}

\begin{document}

\title{From classical to quantum walk via erasure of which-way information}

\author{Miko\l{}aj Lewandowski}   
\affiliation{Faculty of Physics, Adam Mickiewicz University, Umultowska 85, 61-614 Pozna\'n, Poland}

\author{Tomasz Kopyciuk}   
\affiliation{Faculty of Physics, Adam Mickiewicz University, Umultowska 85, 61-614 Pozna\'n, Poland}

\author{Pawe\l{} Kurzy\'nski}   \email{pawel.kurzynski@amu.edu.pl}   
\affiliation{Faculty of Physics, Adam Mickiewicz University, Umultowska 85, 61-614 Pozna\'n, Poland}

\date{\today}


\begin{abstract}

Decoherence transforms a ballistic quantum walk into a diffusive classical random walk. After each step the environment measures the particle's path and the outside world gets to know the which-way information. The relation between the which-way information and the classicality of the spatial distribution is clearly visible in the multi-coin quantum walk in which the information about particle's path is encoded in the multi-coin degree of freedom. The more information is stored on the coins, the more classical and diffusive the spatial distribution is. Here, we propose a generalized version of the quantum eraser scenario that allows for complex erasure strategies. We show that the which-way information can be erased from the coins and the ballistic features in the spatial probability distribution can be recovered.  

\end{abstract}

\maketitle


\section{Introduction}

In a classical random walk a coin is tossed and the outcome determines whether a particle moves one step to the left or right. Such a process generates a diffusive spread and the corresponding spatial probability distribution is Gaussian. After each coin toss the outcome is recorded on some auxiliary system and, if these records are available, the particle's path can be traced back to its initial position. 

A quantum analogue of this process uses quantum coins and quantum particles. The coin toss and the outcome dependent shift leave the system in a superposition state of moving to the right with the coin showing heads and moving to the left with the coin showing tails. The basic feature of any quantum superposition is that there is no information in the entire universe about the exact value of a property whose eigenstates are superposed. Whenever such information is created, as a result of some measurement, the superposition becomes a mixture. However, as long as the state of the system is not measured, the superposition exists and an interference can occur. The interference is an important feature of the quantum walk and is related to its ballistic spreading \cite{Aharonov,Review1,Review2,Review3,Review4}.    

The difference between a superposition and a mixture becomes evident if one adds a decoherence to the quantum walk model \cite{Review2}. Decoherence transforms a ballistic spreading into a classical Gaussian diffusion. During this non-unitary transformation the environment gets to know the particle's path. However, there exists an alternative approach to decoherence in quantum walks. It is a unitary multi-coin quantum walk \cite{MCQW1,MCQW2}, which provides an intuitive explanation of why the knowledge of the particle's path is important. It uses many coins to control the particle's movement and if each step is governed by a different coin, the resulting spatial probability distribution is classical. This is because the particle's path is encoded on coins and each possible path corresponds to a different multi-coin state. As a result, the coin and the spatial degrees of freedom become strongly correlated and the reduced spatial state is a classical mixture at all times.     

The quantum walk is analogous to a sequence of interference experiments. The complex interference pattern is observed after the particle passes through many layers of slits, whose arrangement resembles the classical Galton board. However, if each slit were monitored by some detector, the particle's path would be recorded and the final interference pattern would be lost. In this sense, the multi-coin quantum walk resembles the monitored interference experiment in which the role of the detectors is played by the coins. In this work we show that it is possible to erase the which-way information stored on the coins and to recover the original interference pattern and the ballistic spreading. The idea is based on the standard double-slit quantum eraser scenario \cite{QEraser1,QEraser2}. We propose a version of a multi-coin quantum walk to which a generalized erasure scheme can be applied. It offers various erasure strategies and allows to convert classical diffusive probability distributions into ones exhibiting quantum ballistic features. 

Apart from investigating the relation between classical and quantum dynamics, our work contributes to investigations on measurement-induced effects in quantum walks, which is a developing subfield within the quantum walk studies. It was recently reported that quantum walk dynamics radically changes if the measurement is done in the middle of the evolution \cite{Nitsche}. However, measurement-induced effects date back to the seminal quantum walk paper by Aharonov, Davidovich and Zagury \cite{Aharonov} who showed that a special measurement of the coin degree of freedom can result in a counterintuitive distortion of the spatial probability distribution. This effect was further explored in one- and two-dimensional quantum walks \cite{Rhode}. Another interesting example is an experimental photonic quantum walk with a delayed choice measurement of the coin \cite{DelayedQW}. The authors of this experiment observed that different measurements of polarization, which encoded the state of the coin, lead to different spatial probability distributions. 

We extend the above examples to measurements on much larger Hilbert spaces. This allows for complex measurement scenarios in which the wave function can collapse in various different ways. In particular, we expose the state engineering and the control capabilities of the multi-coin quantum walk model. Due to strong entangling properties of the evolution the manipulation of one set of degrees of freedom can be used to change the state of the other set. Such property can be particularly useful for quantum walk based algorithms \cite{Ambainis}.

\section{Preliminaries}

\subsection*{Discrete-time quantum walks}

\begin{figure}[t]
\includegraphics[scale=0.6]{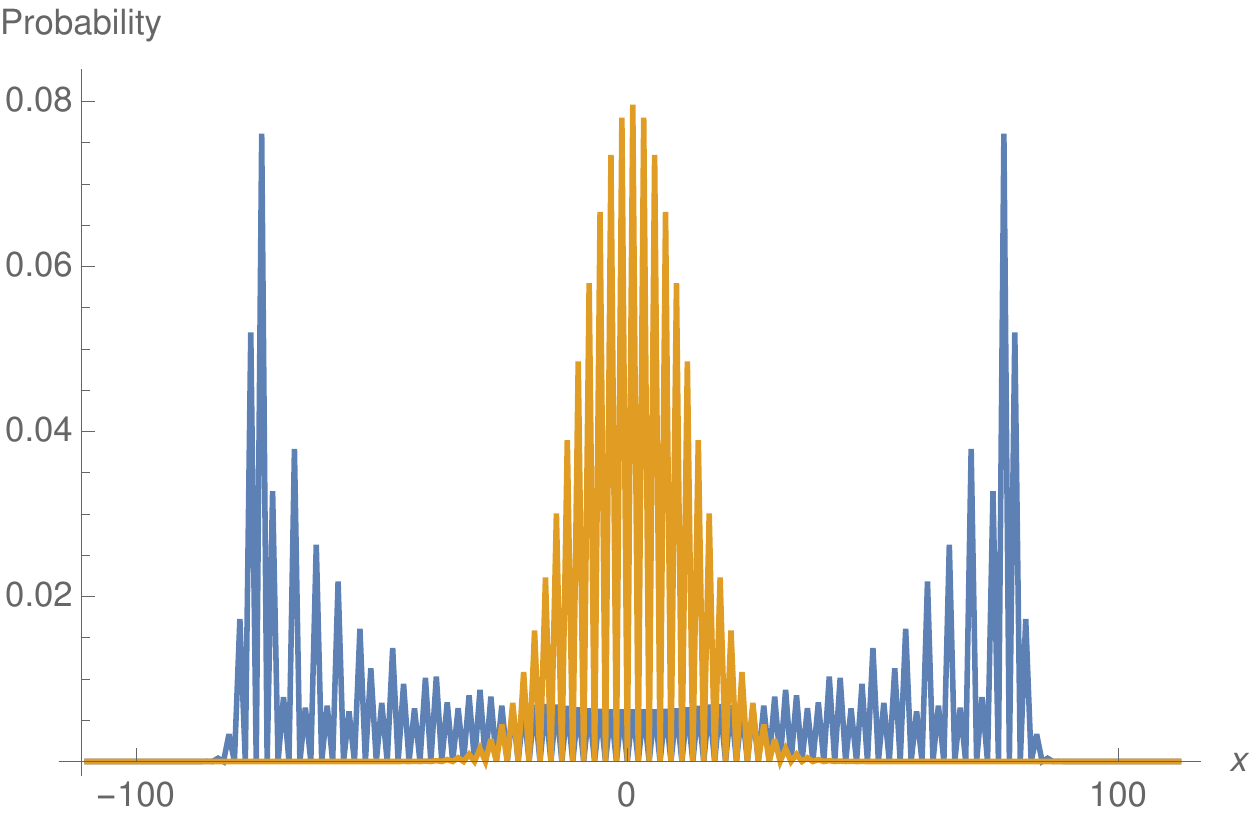}
\caption{Spatial probability distributions of the Hadamard walk (blue) and the classical random walk (orange) after 100 steps. The process starts at $x=0$ and the initial coin state of the Hadamard walk is $\frac{1}{\sqrt{2}}(|0\rangle + i|1\rangle)$. \label{fig1}}
\end{figure}

In this work we consider discrete-time quantum walks (DTQWs) in one dimension. A standard version of this model consists of a particle, whose position is determined by a single integer ($x\in {\mathbb Z}$), and a two-dimensional coin described by a binary variable ($c=0,1$). The state of the system takes form
\begin{equation}
|\psi\rangle = \sum_{x,c} \alpha_{x,c} |x\rangle \otimes |c\rangle.
\end{equation}
The one step of the evolution is governed by
\begin{equation}
U=S(\openone_x \otimes C),
\end{equation}
where $\openone_x$ is the identity operator on the position space, $S$ is the conditional translation operator
\begin{equation}
S|x\rangle\otimes |c\rangle = |x+(-1)^c\rangle \otimes |c\rangle
\end{equation}
and $C$ is a coin 'toss' operator. The operator $C$ is often taken to be the Hadamard transformation $H$
\begin{eqnarray}
H|0\rangle &=& \frac{1}{\sqrt{2}}(|0\rangle + |1\rangle), \\
H|1\rangle &=& \frac{1}{\sqrt{2}}(|0\rangle - |1\rangle),
\end{eqnarray}
therefore the corresponding DTQW is called the Hadamard walk. If the initial state is $|\psi_0\rangle$, the state after $T$ steps is given by
\begin{equation}
|\psi_T\rangle = U^T|\psi_0\rangle.
\end{equation}
The spatial probability distribution corresponding to the above state 
\begin{equation}
p(x,T)=\langle \psi_T | (|x\rangle\langle x|\otimes \openone_c)|\psi_T\rangle,
\end{equation}
where $\openone_c$ is the two-dimensional identity operator on the coin space, is strikingly different than classical Gaussian distribution (see Fig. \ref{fig1}). Apart from its anti-Gaussian shape, the DTQW distribution is ballistic, i.e., its standard deviation is proportional to $T$. 


\subsection*{Multi-coin quantum walks}

Next, we discuss the multi-coin quantum walk (MCQW) that was introduced in \cite{MCQW1,MCQW2}. Here, we focus on a special case for which the number of coins is equal to the number of steps $T$. The state of the system is given by
\begin{equation}
|\psi\rangle = \sum_{x,c_1,\ldots,c_T} \alpha_{x,c_1,\ldots,c_T} |x\rangle \otimes \left(\bigotimes_{j=1}^T |c_j\rangle\right),
\end{equation}
where each $c_j=0,1$. The $i$-th step of the evolution is given by 
\begin{equation}\label{Ui}
U_i=S_i(\openone_x\otimes\openone_c \otimes \ldots \otimes C_i \otimes \ldots \openone_c),
\end{equation}
where $S_i$ is the position shift conditioned on the $i$-th coin
\begin{equation}
S_i |x\rangle \otimes \left(\bigotimes_{j=1}^T |c_j\rangle\right) = |x+(-1)^{c_i}\rangle \otimes \left(\bigotimes_{j=1}^T |c_j\rangle\right)
\end{equation}
and $C_i$ is the coin toss of the $i$-th coin. We choose $C_i = H$ for all $i$. The state after $T$ steps is
\begin{equation}
|\psi_T\rangle = \Pi_{i=1}^T U_{i} |\psi_0\rangle.
\end{equation}  

Due to the fact that for $i\neq j$ we have 
\begin{eqnarray}
& &S_i(\openone_x\otimes\openone_c \otimes \ldots \otimes C_j \otimes \ldots \openone_c)= \nonumber \\
& &(\openone_x\otimes\openone_c \otimes \ldots \otimes C_j \otimes \ldots \openone_c)S_i,
\end{eqnarray}
we can write
\begin{equation}\label{MCQWevolution}
|\psi_T\rangle = \left( \Pi_{i=1}^T S_i \right) \left(\openone_x\otimes H^{\otimes^T}\right)|\psi_0\rangle,
\end{equation}
where we explicitly used the Hadamard transformation for all coins. Therefore, the coin operations can be applied before the particle starts to move. In fact, they can be included in the preparation of the initial state
\begin{equation}
|\phi_0\rangle \equiv \left(\openone_x\otimes H^{\otimes^T}\right)|\psi_0\rangle,
\end{equation}
which gives
\begin{equation}\label{coinless}
|\phi_T\rangle \equiv |\psi_T\rangle = \left( \Pi_{i=1}^T S_i \right)|\phi_0\rangle.
\end{equation}
Note, that because of the above, the particle in the MCQW model can be considered as a kind of automaton capable of moving left or right, depending on the programme which is fed into it. In our case it is a quantum automaton that is fed with a programme encoded on the initial multi-coin state. The programme contains all information about displacements and can be in a superposition.

The initial multi-coin state can be of any form. It could be entangled, which would lead to observable effects in the resulting spatial distribution \cite{MCQWent}. However, in this work we focus on the following separable initial state
\begin{eqnarray}
|\phi_0\rangle &=& \left(\openone_x\otimes H^{\otimes^T}\right)|x=0\rangle \otimes|0\rangle^{\otimes^T} \nonumber \\
&=& |x=0\rangle \otimes \frac{1}{\sqrt{2^T}} \sum_{c_1,\ldots,c_T=0}^1|c_1\ldots c_T\rangle, \label{initcoin}
\end{eqnarray}
where we used the shorthand notation $|c_1\rangle\otimes\ldots\otimes|c_T\rangle \equiv |c_1\ldots c_T\rangle$. 

Next, let us apply the shift operations
\begin{eqnarray}
& &|\phi_T\rangle = \left( \Pi_{i=1}^T S_i \right)|\phi_0\rangle = \label{phiT} \\
& &= \frac{1}{\sqrt{2^T}} \sum_{c_1,\ldots,c_T=0}^1|(-1)^{c_1}+\ldots+(-1)^{c_T}\rangle \otimes |c_1\ldots c_T\rangle. \nonumber
\end{eqnarray}
The spatial probability distribution resulting from $|\phi_T\rangle$ is
\begin{equation}\label{condprob}
p(x,T)=\langle \phi_T | (|x\rangle\langle x|\otimes \openone_c^{\otimes^T})|\phi_T\rangle.
\end{equation}
For (\ref{phiT}) the above turns out to be the classical diffusive Gaussian distribution, because each particle's path is encoded on a different multi-coin state. 


\section{Results}

We consider a measurement of the multi-coin state after $T$ steps. The goal is to erase the which-way information from the coin registers and observe how this affects the spatial probability distribution. In particular, we would like to obtain the DTQW distribution and the one that is as uniform as possible. 


\subsection*{Analysis of $|\phi_T\rangle$}

Let us discuss in details properties of the state (\ref{phiT}). It is entangled between the position and the coins. Note, that although the sum has $2^T$ terms, the corresponding Schmidt rank (in this case the rank of the reduced density matrix of the position/coins degree of freedom) is $T+1$. This is because the position is in the range $x\in [-T,T]$ and only every second position is occupied, i.e., $x\in \{T,T-2,\ldots,-T\}$.

It is useful to represent (\ref{phiT}) in the Schmidt form \cite{NC} 
\begin{equation}\label{finalgamma}
|\phi_T\rangle = \sum_{k=0}^{T} \gamma_k |x=2k-T\rangle\otimes|\Gamma_k\rangle,
\end{equation}
where 
\begin{equation}
\gamma_k =\sqrt{{T\choose{k}}2^{-T}}
\end{equation}
and $|\Gamma_k\rangle$ is the Dicke state \cite{Dicke}
\begin{equation}\label{Dstate}
|\Gamma_k\rangle = \frac{1}{\sqrt{T\choose{k}}}\sum_{i}\sigma_i\left(|1\rangle^{\otimes^k}\otimes |0\rangle^{\otimes^{(T-k)}}\right). 
\end{equation}
In the above $T\choose{k}$ is a binomial coefficient and the sum in (\ref{Dstate}) is taken over all different permutations $\sigma_i$ of the multi-coin state with $k$ ones and $T-k$ zeros. The coins in the Dicke states are multipartite entangled (see for example \cite{DickeEnt}). In addition, the terms in (\ref{Dstate}) correspond to all possible paths the particle can take to get to position $x=2k-T$. Note that
\begin{equation}
\left( \Pi_{i=1}^T S_i \right)|x\rangle\otimes|\Gamma_k\rangle = |x+2k-T\rangle\otimes|\Gamma_k\rangle,
\end{equation}
therefore the initial state (\ref{initcoin}) can be represented as
\begin{equation}\label{initgamma}
|\phi_0\rangle = |x=0\rangle\otimes \sum_{k=0}^T \gamma_k|\Gamma_k\rangle.
\end{equation}

The binomial form of coefficients $\gamma_k$ is the reason why the above state leads to the classical Gaussian distribution. Due to the same reason the state (\ref{finalgamma}) is not maximally entangled. It would be maximally entangled if the coefficients $\gamma_k$ were uniform.  

Since we are also interested in obtaining as uniform spatial distributions as possible, it is useful to introduce the momentum-like basis $\{|m\rangle\}$ 
\begin{eqnarray}
& &|x=2k-T\rangle = \frac{1}{\sqrt{T+1}}\sum_{m=0}^T e^{i\frac{2\pi}{T+1}km}|m\rangle, \\
& &|m\rangle = \frac{1}{\sqrt{T+1}}\sum_{k=0}^T e^{-i\frac{2\pi}{T+1}km}|x=2k-T\rangle.
\end{eqnarray}  
The states $|m\rangle$ correspond to uniform spatial distributions in the region that can be occupied by particle after $T$ steps. Next, we represent the state (\ref{finalgamma}) in the new basis 
\begin{eqnarray}
|\phi_T\rangle &=& \frac{1}{\sqrt{T+1}}\sum_{m=0}^{T} |m\rangle\otimes\left(\sum_{k=0}^T \gamma_k e^{i\frac{2\pi}{T+1}km}|\Gamma_k\rangle\right) \label{finalgammap} \\
&\equiv & \frac{1}{\sqrt{T+1}}\sum_{m=0}^{T} |m\rangle \otimes |G_m\rangle. \nonumber
\end{eqnarray}
The above may look like a maximally entangled state, however the states
\begin{equation}\label{Gp}
|G_m\rangle = \sum_{k=0}^T \gamma_k e^{i\frac{2\pi}{T+1}km}|\Gamma_k\rangle
\end{equation}
are not orthogonal.

Because the state (\ref{finalgamma}) (alternatively (\ref{finalgammap})) is not maximally entangled, the manipulation capabilities of the spatial probability distribution by measurements on the multi-coin state are limited. For example, if one projected the multi-coin state onto $|G_m\rangle$, the spatial state would not be given by $|m\rangle$, but by a superposition $\sum_{m'} \alpha_{m'} |m'\rangle$, where the coefficients $\alpha_{m'}$ are proportional to $\langle G_m|G_m'\rangle$. Such a superposition does not correspond to a ballistic distribution. Therefore, we need to look for more sophisticated erasure strategies to observe a diffusive-to-ballistic transition in our model.


\subsection*{Which-way information erasure}

Imagine that after the MCQW evolution the multi-coin state is measured and the outcome corresponding to the projector $\Pi$ is registered. Due to the quantum nature of the measurement process, the original multi-coin state is altered and the new spatial probability distribution is 
\begin{equation}
p(x,T|\Pi) = \frac{1}{\cal N}\langle \phi_T | (|x\rangle\langle x|\otimes \Pi)|\phi_T\rangle,
\end{equation}
where
\begin{equation}
{\cal N} = \sum_{x=-T}^T \langle \phi_T | (|x\rangle\langle x|\otimes \Pi)|\phi_T\rangle.
\end{equation}
For the state (\ref{finalgamma}) the conditional spatial probability distribution is
\begin{equation}\label{cd2}
p(x=2k-T,T|\Pi) = \frac{1}{\cal N}|\gamma_{k}|^2 \langle \Gamma_{k}|\Pi|\Gamma_{k}\rangle.
\end{equation}
The non-maximal entanglement between the position and the coins allows to manipulate the distribution by factors $\langle\Gamma_{k}|\Pi|\Gamma_{k}\rangle \in [0,1]$.  

Although the projector $\Pi$ is in general defined in the $2^T$-dimensional space, it is useful to confine it to the $(T+1)$-dimensional subspace on which the multi-coin state is supported. Let $\Pi=|\pi\rangle\langle\pi|$ be a rank-1 projector, such that
\begin{equation}\label{pistate}
|\pi\rangle = \sum_{k=0}^T \alpha_k |\Gamma_k\rangle,
\end{equation} 
where $\sum_k |\alpha_k|^2 = 1$. The equation (\ref{cd2}) becomes 
\begin{equation}\label{cd3}
p(x=2k-T,T|\Pi) = \frac{1}{\cal N}|\gamma_{k}|^2 |\alpha_{k}|^2.
\end{equation}
Given a conditional probability distribution $p(x,T|\Pi)$, the coefficients $\alpha_k$ can be expressed in the following form
\begin{equation}\label{coeff}
\alpha_{k} = \frac{\sqrt{{\cal N}p(x=2k-T,T|\Pi)}}{\gamma_{k}}.
\end{equation}


\section*{Recovering DTQW distribution}

Our next goal is to recover the Hadamard walk distribution $p(x,t|\Pi)$ (see Fig. \ref{fig1}). It was evaluated using combinatorial methods in \cite{MeyerComb,AmbComb,CartComb1,CartComb2} as
\begin{equation}\label{Haddist}
p(x=2k-T,T|\Pi)=|\psi_0(2k-T)|^2 + |\psi_1(2k-T)|^2,
\end{equation}
where
\begin{eqnarray}
\psi_0(2k-T)&=&\sum_{j=0}^{\infty} \frac{(-1)^{T-k-j}}{\sqrt{2^T}} {{T-k-1}\choose{j-1}}{{k}\choose{j}}, \\
\psi_1(2k-T)&=&\sum_{j=0}^{\infty} \frac{(-1)^{T-k-j-1}}{\sqrt{2^T}} {{T-k-1}\choose{j}}{{k}\choose{j}}.
\end{eqnarray}  
The summation index goes to infinity, but we use the convention ${x\choose{y}}\equiv 0$ if $y>x$ or $y<0$, therefore the number of non-zero terms is finite. 

The above formula does not work for $k=0,T$, in which case $p(x=\pm T,T|\Pi)=2^{-T}$. The distribution (\ref{Haddist}) corresponds to the DTQW initial state $|x=0\rangle\otimes|0\rangle$. The distribution for the initial state $|x=0\rangle\otimes|1\rangle$ is the same as the one above, but mirror reflected with respect to $x=0$. Finally, the distribution for the initial state  $\frac{1}{\sqrt{2}}|x=0\rangle\otimes(|0\rangle \pm i |1\rangle)$ is an even mixture of the previous two.

\begin{figure}[t]
\includegraphics[scale=0.6]{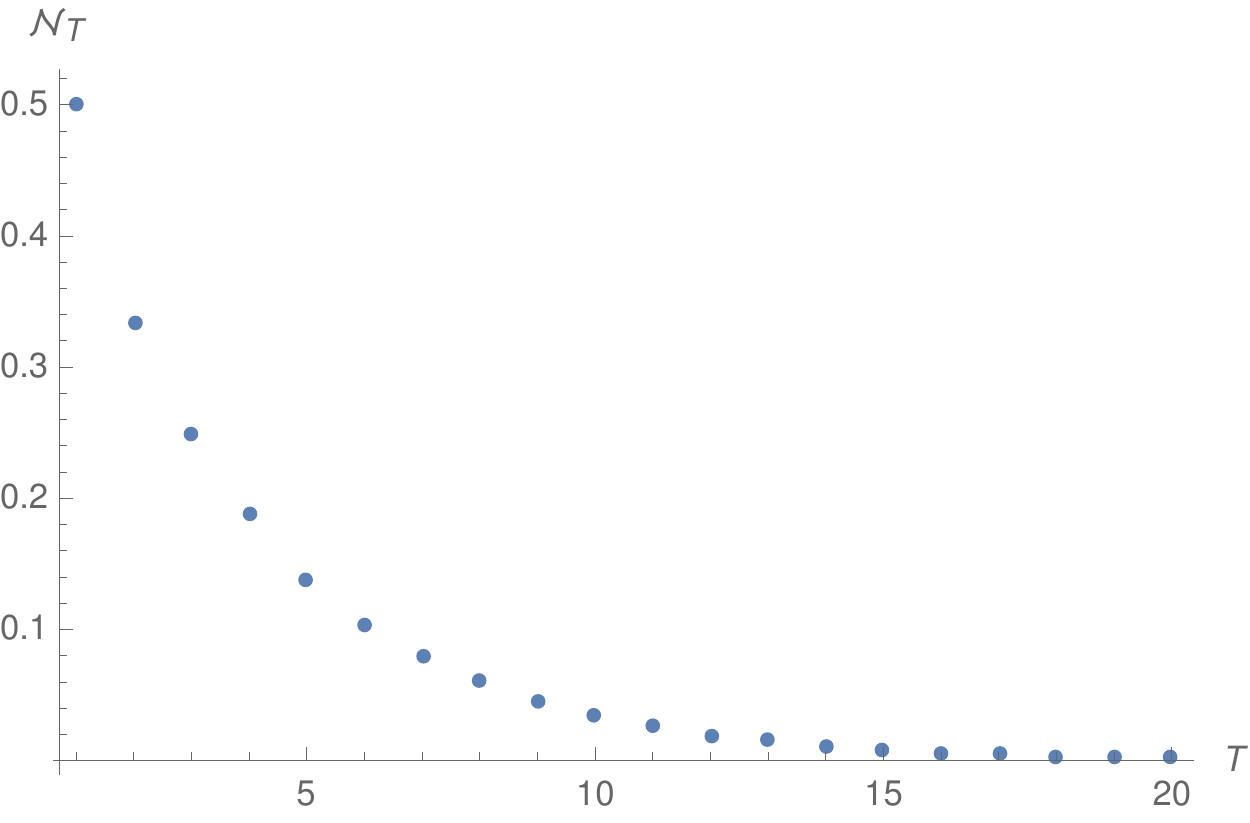}
\caption{First few coefficients ${\cal N}_T$ for the path erasure scenario leading to the Hadamard walk distribution. The value of each coefficient corresponds to the probability of successful projection of the multi-coin state onto the state $|\pi(T)\rangle$.  \label{fig2}}
\end{figure}

The states $|\pi\rangle \equiv |\pi(T)\rangle$, onto which the multi-coin state is projected, can be evaluated using Eqs. (\ref{pistate}) and (\ref{coeff}). For example, the state $|\pi(5)\rangle$ can be represented in the Dicke basis as
\begin{equation}
|\pi(5)\rangle = \sqrt{{\cal N}_5}\left(1, \sqrt{\frac{11}{5}}, \sqrt{\frac{2}{5}}, \sqrt{\frac{2}{5}}, \sqrt{\frac{11}{5}}, 1 \right)^{\cal T},
\end{equation}
where ${\cal N}_5 = 5/36$ and ${\cal T}$ denotes the transposition. The normalization coefficients ${\cal N}_T$ (the probabilities of successful projection) are presented in Fig. \ref{fig2} for the first few values of $T$. 

Let us also discuss one important issue. For the first few steps the classical Gaussian distribution and the Hadamard one are the same. Let us focus on $T=1$. In this case the coefficient ${\cal N}_1 = 1/2$, but there is no point to project the coin onto 
\begin{equation}
|\pi(1)\rangle = \frac{1}{\sqrt{2}}(|\Gamma_0\rangle + |\Gamma_1\rangle)=\frac{1}{\sqrt{2}}(|0\rangle + |1\rangle),
\end{equation}
since the spatial distribution is already in the desired state. It is best to do nothing. This suggests that there might be a better measurement strategy, even for arbitrary $T$.


\subsection*{Uniform distribution}

The Hadamard distribution is quite peculiar. Apart from ballisticity, which is its most celebrated feature, it is quite non-uniform. This results in highly complicated form of $\Pi$. The same is true for any other DTQW distribution. However, if we limit ourselves solely to ballistic behaviour, we can focus on much simpler distributions -- the uniform ones $p(x,T|\Pi)=\frac{1}{T+1}$. 

Let us recall the state (\ref{finalgamma}) and its alternative form (\ref{finalgammap}). The corresponding reduced multi-coin state is given by
\begin{equation}\label{reducedcoin}
\rho_{c}=\sum_{k=0}^T |\gamma_k|^2|\Gamma_k\rangle\langle\Gamma_k|=\frac{1}{T+1}\sum_{m=0}^T |G_m\rangle\langle G_m|.
\end{equation}
Although the states $\{|G_m\rangle\}$ are not orthogonal, they exhibit a particular symmetry. Consider a unitary operator
\begin{equation}
V=\sum_{k=0}^T e^{i\frac{2\pi}{T+1}k}|\Gamma_k\rangle\langle\Gamma_k|
\end{equation}
and its action on states (\ref{Gp})
\begin{equation}
V^n|G_m\rangle = |G_{m+n}\rangle.
\end{equation}
The operator $V$ generates a cyclic permutation of these states.

We aim to obtain a uniform spatial distribution, therefore we need to detect only one of $T+1$ non-orthogonal states $\{|G_m\rangle\}$ in the mixture (\ref{reducedcoin}). However, discrimination between non-orthogonal states cannot be done with perfect efficiency \cite{Helstrom}. Nevertheless, it is possible to use probabilistic methods, which sometimes lead to inconclusive results. Here we are going to use the special case of the maximum confidence quantum measurements developed in \cite{MaxConf}. The authors of this work showed that for a set of (non-orthogonal) pure states $\{|\psi_i\rangle\}$ and the density operator of the form
\begin{equation}
\rho = \sum_i p_i |\psi_i\rangle\langle\psi_i|,
\end{equation}
where $\{p_i\}$ are probabilities, the optimal positive operator valued measure (POVM) to distinguish between these states is given by the set of operators $\{\Pi_i\}$ such that
\begin{equation}
\Pi_i \propto \rho^{-1} |\psi_i\rangle\langle \psi_i| \rho^{-1}.
\end{equation}
In our case 
\begin{equation}
\rho_c^{-1} = \sum_{k=0}^T |\gamma_k|^{-2} |\Gamma_k\rangle\langle \Gamma_k|
\end{equation}
and
\begin{equation}
\rho_c^{-1}|G_m\rangle = \sum_{k=0}^T \gamma_k^{-1} e^{i\frac{2\pi}{T+1}km}|\Gamma_k\rangle \equiv |\tilde{G}_m\rangle,
\end{equation}
therefore the POVM elements are proportional to $|\tilde{G}_m\rangle\langle\tilde{G}_m|$. 

The unnormalized set $\{|\tilde{G}_m\rangle\}$ obeys the same symmetry as $\{|G_m\rangle\}$, i.e., 
\begin{equation}\label{sym}
V^n|\tilde{G}_m\rangle = |\tilde{G}_{m+n}\rangle.
\end{equation}
In addition, it has an important property
\begin{equation}\label{orth}
\langle \tilde{G_m}|G_n\rangle = (T+1)\delta_{m,n},
\end{equation}
where $\delta_{m,n}$ is the Kronecker delta. Therefore, the element corresponding to $|\tilde{G}_m\rangle\langle\tilde{G}_m|$ is registered only if the multi-coin state was $|G_m\rangle$. As a result, the entanglement between the coins and position causes the position state to collapse onto the uniformly distributed state $|m\rangle$. The which-way information erasure is successful.

The set of our POVM elements is given by
\begin{equation}
\Pi_m = \eta_m|\tilde{G}_m\rangle\langle\tilde{G}_m|,
\end{equation}
where $\eta_m$ are non-negative coefficients chosen to satisfy
\begin{equation}
0 \leq \Pi_m \leq \openone,~~\sum_{m=0}^T \Pi_m \leq \openone.
\end{equation}
There is an additional element $\Pi_{?}$ such that
\begin{equation}
0 \leq \Pi_? \leq \openone,~~\Pi_{?} + \sum_{m=0}^T \Pi_m = \openone.
\end{equation}
The procedure fails if the outcome corresponding to $\Pi_{?}$ is registered.

The set $\{\eta_m\}$ constitutes a measurement strategy. In the following part we choose all $\eta_m=\eta$ since the multi-coin state (\ref{reducedcoin}) is an even mixture of $|G_m\rangle$ and we do not favour any of the states $|m\rangle$. The goal is to find the maximal possible value of $\eta$. It is equal to $1/\lambda_{\max}$, where $\lambda_{max}$ is the largest eigenvalue of 
\begin{equation}
\sum_{m=0}^T |\tilde{G}_m\rangle\langle\tilde{G}_m| = \sum_{m,k,k'=0}^T  \gamma_k^{-1}\gamma_{k'}^{-1} e^{i\frac{2\pi}{T+1}(k-k')m}|\Gamma_k\rangle\langle \Gamma_{k'}|.
\end{equation}
Because of symmetry (\ref{sym}) and due to the fact that
\begin{equation}
\sum_{m=0}^T e^{i\frac{2\pi}{T+1}(k-k')m} = (T+1)\delta_{k,k'},
\end{equation}
we get
\begin{equation}
\sum_{m=0}^T |\tilde{G}_m\rangle\langle\tilde{G}_m| = (T+1)\sum_{k=0}^T  \gamma_k^{-2} |\Gamma_k\rangle\langle \Gamma_{k}|.
\end{equation}
The largest eigenvalue corresponds to $k=0$ or $k=T$, for which $\gamma_0^{-2}=
\gamma_T^{-2}=2^T$, hence
\begin{equation}
\eta=\frac{1}{2^T(T+1)}.
\end{equation}
Therefore, using Eq. (\ref{orth}), we find that the probability of detecting state $|G_m\rangle$ is
\begin{equation}
\langle G_m| \Pi_m |G_m\rangle = 2^{-T}.
\end{equation} 

Next, observe that in order to obtain the uniform distribution it is enough to detect any of the states $|G_m\rangle$. As a result, the probability of successful which-way information erasure is
\begin{equation}\label{success}
P_{success}=(T+1)2^{-T}.
\end{equation} 
Note, that this probability of success is optimal. This is because if the conditional distribution $q(x)$ originates from $p(x)$ with probability $P_{success}$, then $P_{success}q(x)\leq p(x)$ for all $x$. Therefore, the optimal probability is given by $P_{success}=\min_x\{p(x)/q(x)\}$. In our case $q(x) = 1/(T+1)$ for all allowed positions and the minimal $p(x)$ is $2^{-T}$, hence the optimal probability is $(T+1)2^{-T}$, which agrees with (\ref{success}).  

The above probability may seem small, however note that the conditional uniform distribution is radically different than the original Gaussian one. It is somehow expected that extreme measurement-induced effects are not occurring with high probability \cite{Aharonov}. Nevertheless, the probability (\ref{success}) is not hopelessly bad from the experimental point of view. Note that many quantum walk implementations are done on heralded single photon sources based on spontaneous parametric down-conversion (SPDC) (see for example \cite{DelayedQW,QWE1,QWE2,QWE3,QWE4,QWE5}). The common SPDC source generates up to $10^6$ entangled photon pairs per second. Therefore, for sufficiently low $T$ one should be able to collect enough data in reasonable time. As a side note, for large $T$ the problem would rather be to construct the MCQW setup than to collect the data.

\begin{figure}[t]
\includegraphics[scale=0.6]{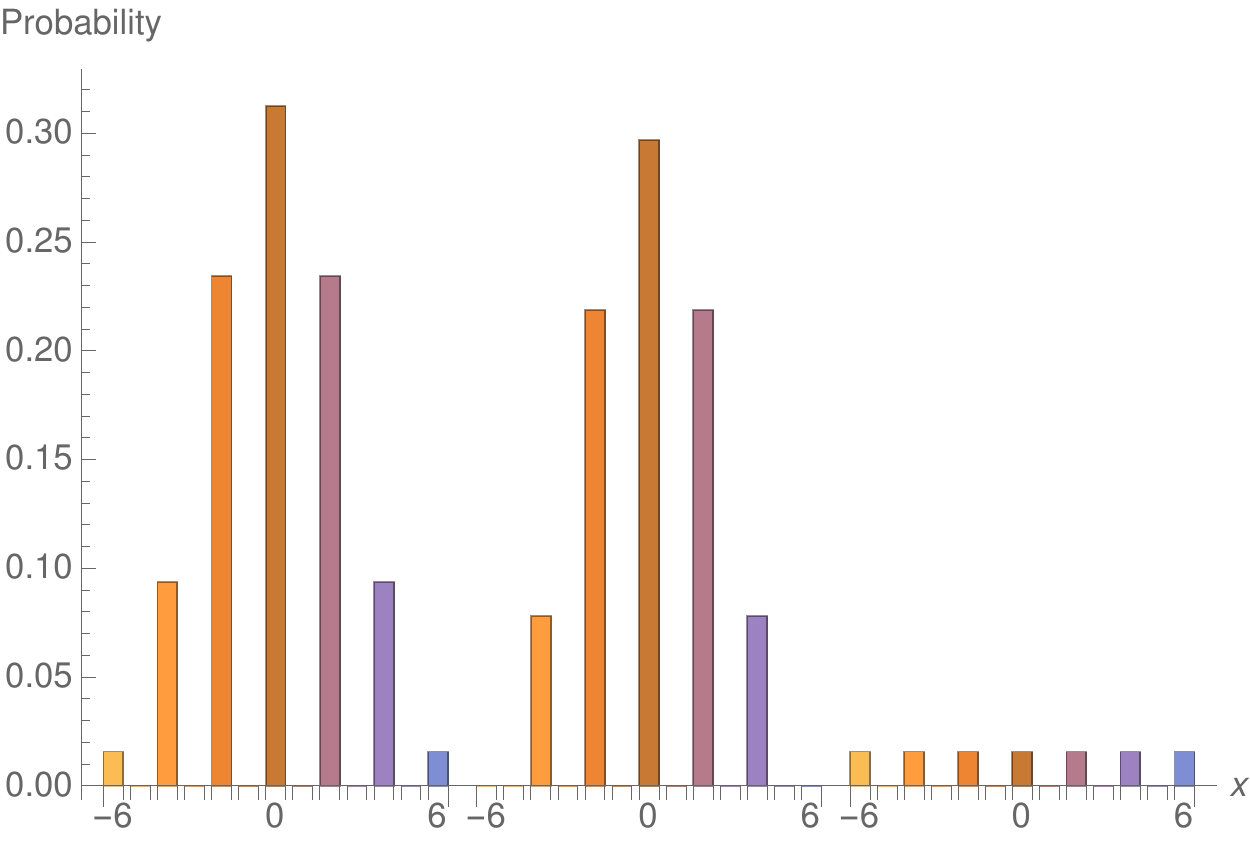}
\caption{Spatial probability distributions for $T=6$. Left -- original Gaussian distribution. Middle -- conditional distribution after registering outcome corresponding to $\Pi_?$. Right -- conditional distribution after registering outcome corresponding to one of the operators from the set $\{\Pi_m\}$. The conditional distributions are not normalized in order to show their relation to the original one. Instead, they are multiplied by $1-P_{success}$ and $P_{success}$, respectively.  \label{fig3}}
\end{figure}

Finally, let us discuss what happens if the outcome corresponding to the element
\begin{equation}
\Pi_? = \openone - \sum_{m=0}^T \Pi_m = \sum_{k=0}^T  \left(1-\frac{k!(T-k)!}{T!}\right) |\Gamma_k\rangle\langle \Gamma_{k}|
\end{equation}
is registered. Since $\Pi_?$ complements the set $\{\Pi_m\}$ to identity, the corresponding spatial distribution needs to complement the conditional uniform distribution to the original Gaussian one. More precisely, just like above, let $p(x)$ be the original Gaussian distribution, $q(x)$ be the conditional uniform one and $r(x)$ be the conditional one corresponding to $\Pi_?$. We have
\begin{equation}
p(x) = P_{success}q(x) + (1-P_{success})r(x),
\end{equation} 
therefore 
\begin{equation}
r(x) = \frac{p(x)-P_{success}q(x)}{1-P_{success}}.
\end{equation}
An example of these three distributions is presented in Fig. \ref{fig3} for $T=6$.


\section{Conclusions}

We proposed a protocol for the erasure of which-way information in the MCQW model. It can change the spatial probability distribution and its importance stems from the fact that the change is caused by a measurement on the auxiliary multi-coin degree of freedom. In particular, we showed that one can obtain two ballistic distributions, the Hadamard walk one and the uniform one, from the diffusive Gaussian distribution. Therefore, our protocol leads to a diffusive-to-classical transition, which can be interpreted as a special type of the classical-to-quantum transition.

Our results show new fundamental connections between classical and quantum walks and they exploit the entangling and manipulation properties of the model. The protocol can be considered a {\it delayed choice measurement} \cite{DelayedQW}, since the choice of whether to measure and what to measure can be done after the particle performed the walk. The erasure of which-way information happens, because the post-measurement state does not contain the information about the path. This is because the measurements used by us are complementary to the basis which encodes the path information.

\section*{Acknowledgements}

This work is supported by the Ministry of Science and Higher Education in Poland (science funding scheme 2016-2017 project no. 0415/IP3/2016/74).



\end{document}